\input{aipcheck}
\documentclass[
    ,final            
  ]
  {aipproc}
\layoutstyle{8x11single}
\usepackage{amsmath,amssymb,slashed}

\begin{document}

\title{Low Scale Left-Right Symmetry and Warm Dark Matter}


\author{Miha Nemev\v sek}{address={ICTP, Strada Costiera 11, 34151 Trieste, Italy\footnote{email: miha@ictp.it}\\JSI, Jamova 39, 1000 Ljubljana, Slovenia}}

\begin{abstract}
We study the scenario of dark matter in the minimal left-right symmetric theory at the TeV scale. The only viable candidate is found to be the lightest right-handed neutrino with a mass of keV. To satisfy the dark matter relic abundance, the relic yield is diluted by late decays of the two heavier neutrinos. We point out that the QCD phase transition temperature coincidences with the typical freeze-out temperature governed by right-handed interactions, which helps to alleviate the problem of overproduction. A careful numerical study reveals a narrow window for the mass of the right-handed gauge boson, within the reach of the LHC.
\end{abstract}

\maketitle

%
%
\section{Introduction}

%
Neutrino mass seems to be the only experimentally established particle physics beyond the SM at the moment. Left-right (LR) symmetric theories~\cite{Pati:1974yy} are an elegant framework of neutrino mass origin and one of the main candidates for physics beyond the Standard Model (SM). They offer an explanation of parity violation at low energies and the discovery of the Higgs boson at the LHC further strengthens the idea of spontaneous symmetry breaking, which lies at the core of LR symmetry. This theory was a progenitor for the see-saw mechanism~\cite{Min}, it predicted a non-zero neutrino mass and related the lightness of left-handed neutrinos to the scale of parity breaking. The minimal LR symmetric model (LRSM) naturally embeds the right-handed (RH) Majorana neutrinos and offers a direct probe of lepton number violation directly at the LHC. The process~\cite{Keung:1983uu} which allows for this direct insight is the Drell-Yan production of a heavy RH gauge boson $W_R$, which decays into a lepton and a heavy Majorana neutrino, that subsequently produces another lepton (or anti-lepton due to Majorana nature) and two jets with no missing energy (see~\cite{Senjanovic:2010nq} for a recent review). It is the high-energy counterpart of the textbook probe of lepton number violation, neutrinoless double beta decay~\cite{Racah:1937qq}. Observation of a signal from this low energy experiment on the other hand could provide a phenomenological motivation for TeV scale LR symmetry~\cite{Tello:2010am, Nemevsek:2011aa}, especially if cosmological data continues to constrain the sum of neutrino masses to small values~\cite{Fogli:2008ig}, disfavouring the light neutrino mass contribution~\cite{Vissani:1999tu}.

%
There exist another hint of new physics, which is the existence of Dark Matter (DM), that may exist in the form of elementary particles. Since the SM fails to provide a stable candidate compatible with the matter energy density observed in the universe, it is a natural question to raise whether one can connect the origin of neutrino mass within the context of LR symmetry to the properties of a DM particle. This is the topic of~\cite{Nemevsek:2012cd}, upon which the present talk is based. The LRSM requires the existence of new electrically neutral states, which might be able to explain the DM puzzle, if the candidate is sufficiently long-lived. For example, the model predicts three RH neutrinos due to LR symmetry and if the lightest RH neutrino is lighter than the pion, its gauge interactions are effectively absent, it decays via Dirac Yukawa couplings only. Since those are governed by small neutrino masses and are small when the RH neutrino is light, the candidate becomes cosmologically stable.

%
A DM candidate with a mass around a keV belongs to the category of warm DM~\cite{Pagels:1981ke, Olive:1981ak}. It provides the same solution for the large scale structure formation as does its heavier cold DM counterpart and it has the advantage of suppressing small scale formations due to free streaming~\cite{Colombi:1995ze}. Therefore, the difficulties of cold DM candidates related to cuspy halo profiles and overpopulated low mass satellite galaxies are alleviated here. A keV RH neutrinos as warm DM has been introduced almost thirty years ago in~\cite{Olive:1981ak, Dodelson:1993je}, where super-weak gauge interactions keep the RH neutrinos in equilibrium. They provide the same picture of thermal production as for the light neutrinos, and this is precisely the case for LR symmetry not far from the TeV scale. The cosmological constraints on such a scenario, in particular a potential over abundance of sterile neutrinos was considered in~\cite{Olive:1981ak} and seemingly prevents the possibility of having a DM candidate in the LRSM. A beautiful solution to this problem was presented in~\cite{Scherrer:1984fd}, where the authors show how a massive long-lived particle affects the evolution of the universe and may produce a significant amount of entropy to open up new parts of parameter space, for example in the case of a heavy RH neutrino~\cite{Asaka:2006ek}.

%
The idea of late entropy production has been applied to generic gauge extensions of the SM in~\cite{Bezrukov:2009th}. There, a lower limit on the mass of $W_R$ around 10\,--\,16 TeV was obtained, which is outside the possible reach of the LHC. The intriguing possibility of testing warm DM at the LHC motivated us~\cite{Nemevsek:2012cd} to reconsider their result in more detail. While the basic picture of~\cite{Bezrukov:2009th} is correct, we find that the exact lower limit is a more complex issue and a possible window for warm DM is revealed in the few TeV region.

%
The main reason behind the possible existence of a light $W_R$ window is related to a simple fact that the freeze-out temperature of TeV scale gauge interactions is on the same order as the QCD phase transition. A dramatic change in the number of relativistic degrees of freedom allows to separate the relic abundances of different species of RH neutrinos. In order to achieve this, the interaction strength of RH neutrinos should be sufficiently different, which can be achieved within a particular portion of parameter space. A particular flavour structure is therefore expected, since both, the production and decays of RH neutrinos are governed by gauge interactions when the scale is low. In fact, a peculiar mass spectrum emerges as a consequence of the requirement to have an acceptable amount of DM in the universe.

%
The basic idea of this scenario in the limit when the LR scale is very large have been studied extensively in the literature. In that limit, the LRSM crosses over to the so-called $\nu$MSM picture~\cite{Asaka:2005an, Asaka:2006ek, Kusenko:2009up} (see also~\cite{Abazajian:2012ys} for a recent white paper on sterile neutrinos). In this case, the role of gauge interactions is taken over by the Yukawa interactions and an appropriate amount of DM can be obtained in a suitable portion of parameter space. In this sense, the LR theory itself does not have a problem to accommodate DM, the point of interest is how it may be distinguished from the $\nu$MSM case, therefore a precise lower bound on the LR scale is particularly important. Our results show that there exist a possibility of a fairly light $W_R$ with rich phenomenology related to LHC physics and other low energy processes, such as neutrinoless double beta decay.

%
%
\section{The minimal left-right symmetric model}

Left-right symmetric theories~\cite{Pati:1974yy} are based on the parity symmetric gauge group $SU(2)_L \times SU(2)_R \times U(1)_{B-L}$ (suppressing color), augmented by a LR symmetry, which is either parity or charge conjugation. Fermions are contained in LR symmetric representations
\begin{equation}
	Q_{L,R} = \begin{pmatrix} u \\ d \end{pmatrix}_{L,R}, \quad L_{L,R} = \begin{pmatrix} \nu \\ \ell \end{pmatrix}_{L,R},
\end{equation}
while the Higgs sector of the minimal model~\cite{Min} consists of a bidoublet $\Phi = \left( 2_L, 2_R, 0_{B-L} \right)$ and two triplets, $\Delta_L = \left(3_L, 1_R, 2_{B-L} \right)$ and $\Delta_R = \left(1_L, 3_R, 2_{B-L} \right)$
\begin{equation}  \label{ds32}
  \Phi = 
  \begin{pmatrix} \phi_1^0 & \phi_2^ + 
  \\
  \phi_1^- & \phi_2^0 \end{pmatrix}\, ,
  \quad
  \Delta_{L, R} = 
  \begin{pmatrix} \Delta^+ /\sqrt{2} & \Delta^{++} 
  \\
  \Delta^0 & -\Delta^{+}/\sqrt{2}
  \end{pmatrix}_{L,R} \, .
\end{equation}
The model is completely LR symmetric and parity is broken spontaneously by the vacuum expectation value(vev) of $\Delta_R$ ($v_R$), which is on the order of TeV and provides mass to the heavy gauge bosons 
\begin{equation}
	M_{W_R} = g \, v_R, \quad M_{Z_{LR}} \simeq \sqrt{3} \, M_{W_R}.
\end{equation}
This relation between the masses of the gauge bosons turns out to be of particular significance in this work. 


The final stage of symmetry breaking down to the SM is characterised by the following vevs $\langle \Phi \rangle = \text{diag}\left( v_1, v_2 \right)$, $\langle \Delta^0_{L,R}\rangle = v_{L,R}$, in a hierarchical order $v_L^2 \ll v^2 = v_1^2 + v_2^2 \ll v_R^2$, with $v = 245 \text{ GeV}$. After the first stage of breaking, heavy neutrinos get their Majorana masses proportional to $v_R$. The Dirac Yukawa terms together with $v_{1,2}$ provide the mixing between heavy and light neutrinos in the so-called type I see-saw. A second term is present as well, which is directly proportional to $v_L$ and gives a direct Majorana mass for the light neutrinos~\cite{Mohapatra:1980yp, Magg:1980ut}. Both terms, together with neutrino mass, vanish when the LR scale is very large. For a recent summary of the main features of LRSM and a detailed study of constraints from meson mixing and other CP violating observables, see~\cite{Zhang:2007da, Maiezza:2010ic, Blanke:2011ry}.

The gauge interactions which govern the thermal production of fermions in LRSM are governed by the following Lagrangian
\begin{align}
 {\mathcal L}_{CC} &= \frac{g}{\sqrt{2}} W_R^\mu \left[
  \begin{pmatrix} \overline{N_1} &\overline{N_2} &\overline{N_3} \end{pmatrix}_R
  \mathbf{V}_\ell^{R\dagger} \gamma_\mu \begin{pmatrix}e \\ \mu \\ \tau \end{pmatrix}_{\!\!R} + \begin{pmatrix} 
 \overline{u} &\overline{c} &\overline{t} \end{pmatrix}_R
  \mathbf{V}_q^R \gamma_\mu \begin{pmatrix}d \\ s \\ b \end{pmatrix}_{\!\!R} \right]+\text{h.c.} \,, 
  \\
  {\mathcal L}_{NC} &= \frac{g}{\sqrt{1 - \tan^2 \theta_W}} Z^\mu_{LR} \bar f \gamma_\mu \left[ T_{3R} + \tan^2\theta_W (T_{3L} - Q) \right]f + \frac{g m_N}{2M_{W_R}} \Delta_R^0 N N, \label{neutralcurrent}
\end{align}
where the equality of the gauge couplings is dictated by LR symmetry. The mixing matrices $\mathbf{V}_{\ell}^R$ and $\mathbf{V}_q^R$ which determine the flavour dynamics in the charged current are the right handed analogs of the PMNS and the CKM matrix. The quark mixings in the RH sector are either very close or equal to the left-handed ones, depending on the choice of parity~\cite{Maiezza:2010ic}. These equalities, together with $g_R = g_L$ need not be exact, a small variation due to renormalisation group equation running might be present if parity were broken at a high scale, however this does not affect our conclusions.

Besides being heavier than $W_R$, the neutral $Z_{LR}$ also couples more weakly to the RH neutrinos, as seen form Eq.~\eqref{neutralcurrent}, which implies a higher freeze-out temperature $T_f$ for a state which couples predominantly to it. A small lift in $T_f$ may have a significant impact on the abundance if number of degrees of freedom change rapidly, as happens to be the case around the QCD phase transition.

The LRSM contains a number of states whose masses play a significant role in the dynamics of the early universe, therefore current experimental bounds on particular states should be reviewed. There is a long history of limits derived from precision studies, such as the mixing of neutral kaons~\cite{Beall:1981ze}, which has been revisited and a lower bound was set $M_{W_R} > 2.5\,$--\,$4 \text{ TeV}$ in~\cite{Maiezza:2010ic, Zhang:2007fn, Zhang:2007da, Bertolini:2012pu}, depending on the choice of the LR symmetry, charge conjugation or parity, respectively. It is remarkable that the LHC has finally overtaken this limit and started to probe the LRSM directly at colliders~\cite{CMSATLASKS} searching for the KS reaction~\cite{Keung:1983uu}. The searches for other states in the LRSM have also been updated and limits have been improved substantially, as can be seen in Table 1. The most stringent bound at the moment applies to the mass of the second Higgs doublet, which should be heavier than about 10 TeV due to its tree-level flavour changing couplings. This has recently been studied in great detail~\cite{Blanke:2011ry}.

The only states resisting to searches at the LHC are light RH neutrinos and the neutral component of the right-handed triplet, both singlets under the SM gauge group. Their role as a DM candidate is the subject of the following section.

\begin{table}
\centering
\begin{tabular}{ccrcc}
	\hline
	Particle & Final state & Lower limit $   $ & Collaboration & Comments
	\\ \hline 
	$W_R$ & $jj$ & 1.5 TeV 	& CMS~\cite{CMSjj} & independent of $N$ mass
	\\
	$W_R$ & $e/\mu + N$ &	2.5	TeV 	& CMS~\cite{CMSWR} & light $N$ (missing energy)
	\\
	$W_R$ & $\ell \ell j j$  & $\lesssim 2.5$ TeV	& ATLAS, CMS~\cite{CMSATLASKS} & heavy Majorana $N$~\cite{Nemevsek:2011hz}
	\\
	$Z_{LR}$ & $e^+ e^-/\mu^+ \mu^-$ & $\sim 2$ TeV  & ATLAS~\cite{ATLASZLR} & see~\cite{Langacker:2009su}
	\\
	$Z_{LR}$ & $e^+ e^-$ & $\sim 3$ TeV  & LEP~\cite{LEPZLR} & indirect, see~\cite{Carena:2004xs, Cacciapaglia:2006pk}
	\\
	$\Delta_L^{++}$ & $\ell_i^+\ell_j^+$ & 100-459 GeV & ATLAS, CMS~\cite{Aad:2012cg, DeltaL++}	
	 & spectrum dependent~\cite{Melfo:2011nx}
	\\
	$\Delta_L^{+}$ &  $\slashed{E}_T+j$ & 70-90 GeV & LEP~\cite{Abbiendi:2003ji} & chargino search~\cite{Pierce:2007ut}
	\\	
	$\Delta_L^{0}$ &   & 45 GeV & LEP~\cite{LEPZLR} & $Z$-boson width
	\\	
	$\Delta_R^{++}$ &$\ell_i^+\ell_j^+$ & 113-251 GeV & ATLAS~\cite{Aad:2012cg}, CDF
	\cite{Abazov:2011xx} & flavor dependent
	\\
	\hline
\end{tabular}
\caption{A summary of limits on the mass scales of the particles in LRSM from collider searches. 
\label{tabBound}}
\end{table}

%
%
\section{A Blueprint for Warm Dark Matter}

Here we present a simplified analytical insight into the picture of warm DM for a LRSM in the TeV region. We consider all the possible candidates and single out the RH neutrino as the only option. We study its thermal production in the early universe and estimate its relic (over)abundance. We proceed to analyse the amount of dilution generated by late decays of the remaining two RH neutrinos. This fixes the flavour structure and predicts the masses of all heavy neutrinos.

\subsection{Candidates for DM}

A viable DM candidate should be electrically neutral and cosmologically stable. LRSM contains a number of neutral states, but not all of them can be sufficiently long-lived. The neutral component of the heavy Higgs from the bi-doublet and the neutral state in $\Delta_L$ are both heavy and decay within collider timescales.

A somewhat curious possibility is the neutral component of $\Delta_R$, which is a singlet under the SM group and therefore could be light. Similarly to the SM Higgs boson, it can decay to two photons via the gauge boson loop with an appreciable rate, even when the LR scale is fairly high
\begin{equation}\label{deltalife}
  \Gamma_{\Delta^0_R\to \gamma\gamma} \simeq \frac{49}{8 \pi} \left( \frac{\alpha}{4\pi} \right)^2 
  \left( \frac{M_{W}}{M_{W_R}} \right)^2  \frac{G_F}{\sqrt{2}} m^3_{\Delta} \simeq 10^{-50} \text{ GeV}
  \left(\frac{m_{\Delta}}{\rm keV} \right)^3 \left( \frac{10^{12} \text{ GeV}}{M_{W_R}} \right)^{2}.
\end{equation}
Here, the factor of $49$ comes from the loop function and the typically subdominant contribution from charged scalars has been neglected. Such radiative decays of long-lived particles produce gamma ray lines and there are severe astrophysical restrictions on the decay rate $\Gamma \lesssim 10^{-50}\,\text{ GeV}$~\cite{Abazajian:2001vt}. Even for a very light $\Delta_R^0$ with a mass of a few keV, the LR scale would have to be extremely high, around $10^{12} \text{ GeV}$, quite an unlikely scenario. While this might be of interest for $SO(10)$ theories, where this is a typical intermediate scale~\cite{Rizzo:1981su}, we focus here on a TeV scale realization and therefore do not pursue the case of $\Delta_R$ as DM.

The only remaining candidate is therefore the lightest RH neutrino, $N_1$. Once lighter than all the other states which couple to $W_R$, it can decay via Yukawa couplings only. Although the fastest decay channel is the tree level decay such as $N \to 3 \nu$ via the $Z$ boson, the most stringent constraint comes from the loop amplitude and the radiative decay $N \to \nu \gamma$. Similar to the case above, the astrophysical X-ray constraints on the emitted  monochromatic photon place a limit on the mixing between light and heavy neutrinos~\cite{Smirnov:2006bu, Boyarsky:2009ix} 
\begin{equation}\label{theta}
  \theta_1^2 < \left(1.8-3.1\right)\times 10^{-5} \left( \frac{\text{keV}}{m_{N_1}} \right)^{5}.
\end{equation}
The resulting lifetime for a mass of around keV becomes large enough for $N_1$ to play the role of DM and we stick to it for the rest of this work. It is noteworthy that there is a lower bound on the allowed mixing, even when the mass of the lightest neutrino is very small. This is due to the symmetric properties of the Dirac mass matrix, which allow to compute the Dirac Yukawa matrix once the mass and mixings of heavy and light neutrinos are given~\cite{future}.

Before moving on to the thermal production, some preliminary remarks regarding the mass of $N_1$ and its Dirac couplings are in order. As we already mentioned in the introduction, the main problem we will face in the TeV-scale LRSM is the thermal overproduction of DM~\cite{Olive:1981ak, Bezrukov:2009th} due to right-handed gauge interactions. Since the mass energy density is proportional to the mass of $N_1$, the heavier it is, the more severe this problem becomes and the question of a lower bound becomes relevant. A low mass of $N_1$ should be consistent with the see-saw formula and the general structure of the LRSM. For a $W_R$ in the few TeV range, it turns out that there exist radiative corrections on the Dirac mass~\cite{Branco:1978bz}, which are on the order of a few eV. From the see-saw formula this in turn implies that the lightest RH neutrino should be heavier than around keV or so. 

There exist cosmological lower limits on the DM mass. A conservative bound is derived by the phase space density considerations applied to compact objects~\cite{Tremaine:1979we}. The limit is also around a keV and we will revisit it at the end of the paper. Another important process which further restricts the allowed couplings of $N_1$ is the supernovae cooling. As was pointed out in~\cite{Raffelt:1987yt}, weakly coupled species lighter than 10\,MeV may carry away a significant amount of energy and speed up the cooling. In the context of the LRSM, this implies the following bound on the electron component of $N_1$: $|V^R_{e1}| < (M_{W_R}/ 23\text{ TeV})^2$~\cite{Barbieri:1988av}, which has to be roughly below 1\%, if LR symmetry were close to the TeV scale. This limit poses a constraint on the see-saw content of the model. For example in the case of type II dominance the flavour is fixed $\mathbf{V}_\ell^R \simeq \mathbf{V}_\ell^L$ and with recent data on non-zero $\theta_{13}$~\cite{GonzalezGarcia:2012sz}, we get a bound $M_{W_R} > 21(10) \text{ TeV}$ for the normal (inverted) hierarchy. Conversely, a low scale of LR symmetry with a keV warm DM requires sub-dominance of the type II contribution.

%
\subsection{Thermal freeze-out and relic abundance}

The presence of right-handed gauge interactions keep the RH neutrinos in thermal equilibrium due to scattering with the SM particles. This thermal production is similar to that of light neutrinos and dominates other interactions. The Yukawa couplings themselves are too small to bring them in equilibrium above the electroweak scale and after spontaneous breaking occurs, the mixing between heavy and light neutrinos is further suppressed by matter effects~\cite{Barbieri:1990vx}. Therefore the role of Yukawa interaction can be neglected~\cite{Boyarsky:2009ix}. Non-thermal contribution due to accumulated oscillations might be present~\cite{Dodelson:1993je, Shi:1998km} but plays a minor role compared to thermal processes within LRSM.

One can estimate the number density of RH neutrinos produced in the early universe in analogy to the case of light neutrinos. Due to weaker interactions, we expect RH neutrinos to decouple from the plasma earlier, therefore $T_f$ is expected to be much above MeV. The annihilation rate of $N$ is suppressed by the large mass of heavy gauge bosons and the out-of-equilibrium condition $\Gamma=H$ now gives the following condition in the radiation dominated universe~\cite{Kolb:1990vq1}
\begin{equation}
	G_F^2 \left( \frac{M_W}{M_{W_R}} \right)^4 T_f^5 \simeq  \sqrt{g_*(T_f)} \frac{T_f^2}{M_{\rm p}} .
\end{equation}
Since we are focusing on a low scale of LR symmetry, the freeze-out temperature is estimated
\begin{equation} \label{eqTfEstimate}
	T_{f} \simeq 400\, {\rm MeV} \left(\frac{g_{*}(T_f)}{70}\right)^{1/6} \left( \frac{M_{W_R}}{5\,{\rm TeV}} \right)^{4/3}.
\end{equation}
As suspected, the temperature is significantly higher than that for the light neutrinos and any RH neutrino with a mass below few hundred MeV will be copiously produced. The resulting number density for such fairly light states is approximately
\begin{equation}\label{Nyield}
Y_{N} \equiv \frac{n_{N}}{s} \simeq \frac{135 \, \zeta(3)}{4 \pi^4 \, g_*(T_{f})}.
\end{equation}
Notice that the yield $Y_N$ is inversely proportional to the number of relativistic degrees of freedom, which grows when the freeze-out temperature increases. Therefore, we expect to have less over-abundance of DM when $N_1$ freeze-out is governed by weaker interactions, e.g. by neutral currents only. This happens naturally if it couples predominantly to the $\tau$, in which case the charged interactions cease around $T\sim m_\tau$ due to the Boltzmann suppression which starts to eliminate $\tau$ from the plasma.

The final estimate for the present-day relic abundance can be estimated using the measured values for entropy density $s=2889.2\,{\rm cm}^{-3}$, critical density $\rho_c=1.05368\times10^{-5} h^2\,{\rm GeV/cm^3}$ and $h=0.7$ and is
\begin{equation} \label{dmrelic}
  \Omega_{N_1} = \frac{Y_{N_1} m_{N_1} s}{\rho_c} \simeq 3.3\times \left( \frac{m_{N_1}}{1\,{\rm keV}} \right) \left( \frac{70}{g_{*}(T_{f1})} \right).
\end{equation}
If this were the final result, the amount of DM produced in this way would over-close the universe by at least a factor of 3. The most recent relic abundance of DM observed in~\cite{Komatsu:2008hk}
\begin{equation}\label{pdgrelic}
  \Omega_{\text{DM}} = 0.228 \pm 0.039 \ , 
\end{equation}
at $3\sigma$ confidence level reveals that the problem is even more severe, the amount of DM is too large by roughly a factor of 12. Notice that this problem can not be solved by increasing $g_*$, for it would require about 1000 additional degrees of freedom, not present in the LRSM.

%
\paragraph{Entropy release}

A way out is to dilute the abundance of $N_1$ by entropy release of a massive particle. If the decay products thermalise and reheat the plasma, this process reduces the DM number density. If the amount of entropy produced is sufficient, an appropriate amount of DM density can be obtained in this way. Of course, in order for this to work, the final state of the decaying particle should not contain a significant amount of DM.

A simple and fairly accurate estimate of the dilution factor can be derived in the sudden decay approximation~\cite{Scherrer:1984fd}. At a given moment corresponding to the lifetime of the decaying particle, the universe is dominated by this matter component. In this approximation, we assume all of its energy content is immediately transferred to radiation at $\tau = H^{-1}$, which gets reheated to the temperature $T_r$
\begin{equation}
  T_{r} \simeq 0.78 \, g_*(T_{r})^{-1/4} \sqrt{\Gamma M_{\rm p}} \simeq 1.22\,{\rm MeV} \left( \frac{1\,{\rm sec}}{\tau} \right)^{1/2}.
\end{equation}
Simply by energy conservation $m n_{N}(\tau_{N}) \equiv m Y s =\rho_{R}(T_r)$, a relation between the amount of energy before and after the decay can be obtained
\begin{equation}
  m  Y s_{before} = \frac{3}{4} s_{after}T_r,  
 \end{equation}
which gives the following estimate for the dilution factor
\begin{equation}\label{eqDilutionS}
  \mathcal{S} \equiv \frac{S_\text{after}}{S_{\text{before}}} \simeq  \frac{s_\text{after}}{s_{\text{before}}}\simeq 1.8 \left(g_*(T_r)\right)^{1/4} \frac{Y \, m}{\sqrt{\Gamma M_{\rm p}}} \ ,
\end{equation}
where $M_{\rm p} = 1.2 \times 10^{19} \, $GeV is the Planck scale. Assuming the sudden decay happens when the DM abundance is already fixed, the relic density calculated in Eq.~(\ref{dmrelic}) will be reduced by the dilution factor. For a reheating temperature $T_r$ around MeV we have 
\begin{equation} \label{dilutedmrelic}
  \Omega_{N_1} \simeq (0.228 + 0.039) \left( \frac{m_{N_1}}{1\,{\rm keV}} \right) \left( \frac{1.85\,{\rm GeV}}{m} \right) \left( \frac{1\,{\rm sec}}{\tau} \right)^{1/2} \left( \frac{g_{*}(T_{f2,3})}{g_{*}(T_{f1})} \right).
\end{equation}

Let us see which candidates could play the role of the diluter in the LRSM. In order to have a sufficient amount of DM, a dilution factor on the order of 10 is needed. As seen from Eq.~\eqref{dilutedmrelic}, this could be achieved with a particle of mass around a GeV with a lifetime around a second. This lifetime should not be substantially longer in order to start Big Bang Nucleosynthesis (BBN) with an appropriate proton-neutron number ratio, reheating temperature should be larger than about MeV, corresponding to $\tau \lesssim \mathcal{O}(1)\,$second.

As seen above, the neutral component of $\Delta_R$ decays fairly rapidly and would be lighter that MeV if it were to live up to a second. Therefore its role in entropy production is subdominant and the only candidates are the two remaining RH neutrinos $N_{2,3}$.

Now we see the intrinsic problem of the low scale LRSM, which prevents a sufficient entropy release. Once we decide on a TeV scale for the gauge boson mass, the freeze-out temperature is around a hundred MeV (Eq.~\eqref{eqTfEstimate}). Therefore, a particle with a mass of around GeV, which seems perfect for the job, will already be gone from the plasma due to the Boltzmann suppression and it seems there is no way out and a lower bound of 10--16 TeV would emerge~\cite{Bezrukov:2009th}. 

\begin{figure}[t!]
\centering
\includegraphics[width=7cm]{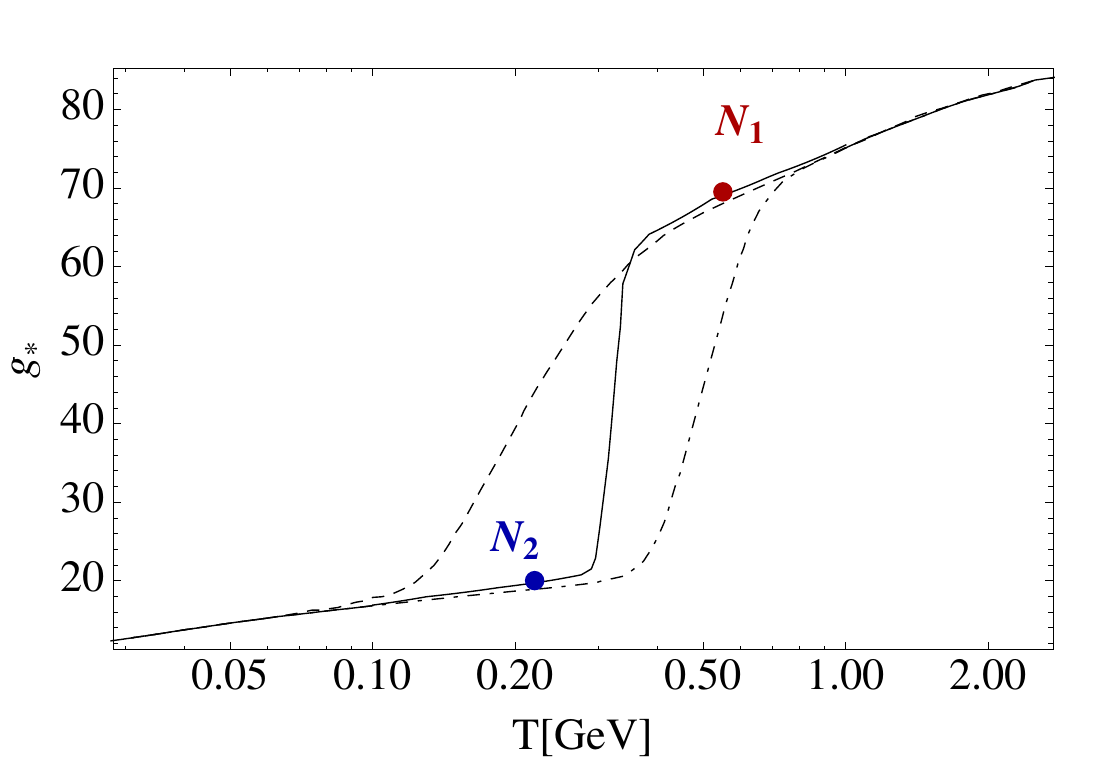}
\includegraphics[width=7cm]{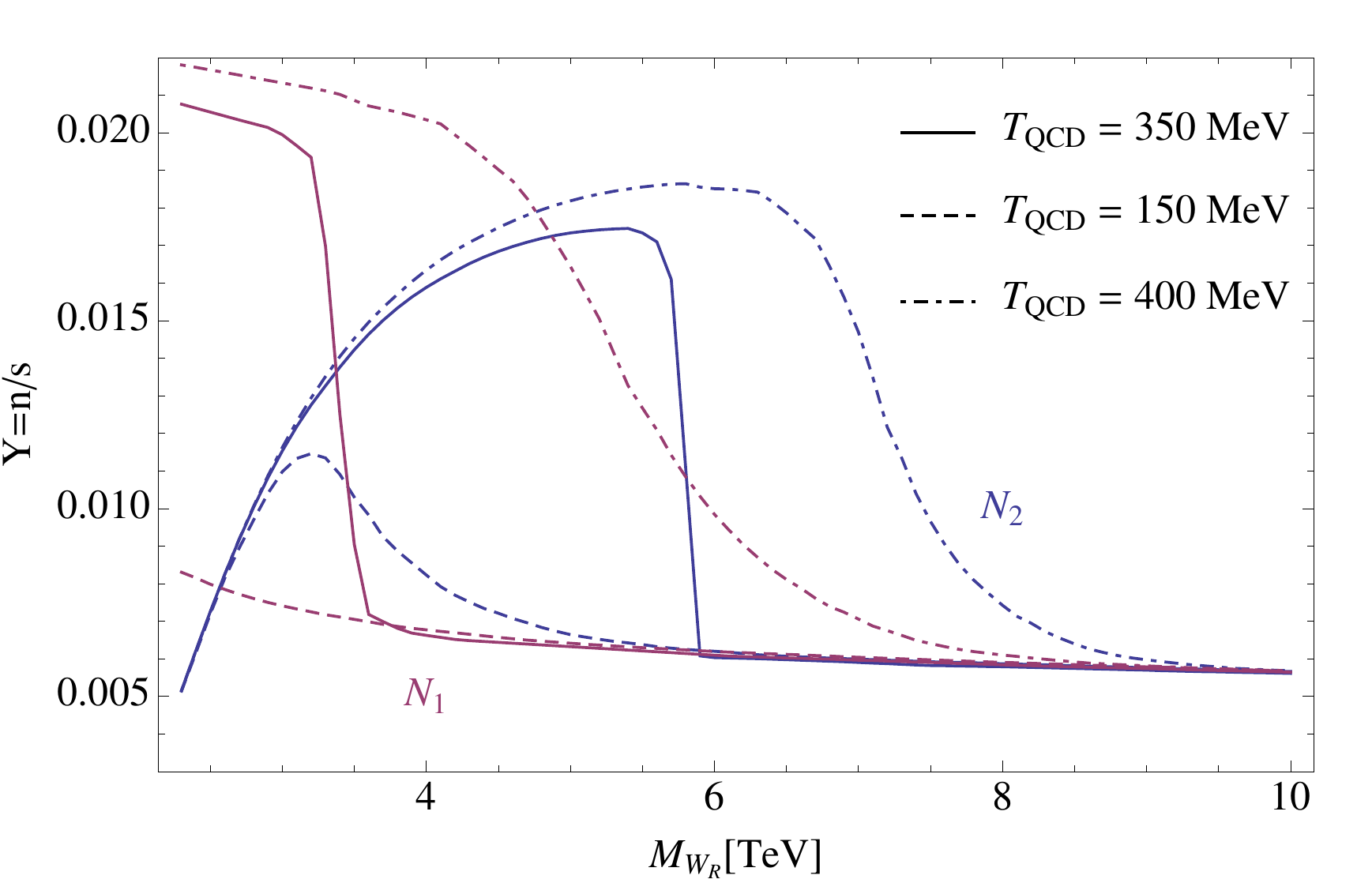}
\caption{Left: A change in the evolution of $g_{*S}$ around the QCD phase transition temperature separates the freeze-out temperatures of $N_1$ and $N_2$ (taken from~\cite{Srednicki:1988ce}). Right: Yields of $N_{1,2}$ at freeze-out is shown in magenta (blue), depending on the mass of $M_{W_R}$, see~\cite{Nemevsek:2012cd} for details.}
\label{figgSYN}
\end{figure}

All may not be lost, for the presence of the QCD phase transition may allow to separate $T_f$ of the two components and reduce the amount of DM produced while preserving the abundance of diluters. This happens when the DM component couples predominantly to $\tau$ and once the temperature drops below $m_\tau$, it is gone from the plasma and the only interactions remaining are those governed by $Z_{LR}$. Due to its heavy mass and reduced couplings to $N$'s, these interactions are weaker and the freeze-out temperature is higher. If the $W_R$ scale is around TeV, this will separate the two components as seen in Fig.~\ref{figgSYN}.

The relative number density between two relativistic RH neutrinos becomes
\begin{equation}\label{YNratio}
  \frac{Y_{N_2}}{Y_{N_1}} \sim \frac{g_{*}(T_{f1})}{g_{*}(T_{f2})},
\end{equation}
which can be as large as $3-4$ for $M_{W_R}\lesssim10 \text{ TeV}$. This is a simplified estimate with a single diluter. A detailed numerical treatment for all RH neutrinos using Boltzmann equations with complete thermally averaged interaction rates has been performed in~\cite{Nemevsek:2012cd}, where the contribution of both RH acting as diluters is treated consistently.

\paragraph{Neutrino decays and mass spectrum}

In order for the diluter's lifetime to reach an order of a second, its mass and couplings have to take on a particular form. The fastest decay rate for a heavy RH neutrino is into a lepton and two jets. When their mass is smaller, the only gauge mediated channel remaining is the pion final state with a lifetime~\cite{Nemevsek:2012cd}
\begin{equation} \label{eqNlPi}
  \tau(N_i\to\ell\pi) = 1 \text{ sec} \left(\frac{m_{N}}{250\,{\rm MeV}}\right)^{-3} \left( \frac{M_{W_R}}{5\,{\rm TeV}} \right)^{4} \left( \frac{0.002}{f(x_\ell, x_\pi)} \right),
\end{equation}
where $f(x_\ell, x_\pi) =\left[ (1-x_\ell^2)^2 - x_\pi^2(1+x_\ell^2) \right] \left[ \left( 1-(x_\pi+x_\ell)^2 \right)\left( 1-(x_\pi-x_\ell)^2 \right) \right]^{1/2}$ and $x_{\pi, \ell}=m_{\pi, \ell}/m_N$. A lifetime on the order of a second narrows down the diluters mass to
\begin{equation} \label{eqmN2Pi}
  m_{N} \approx m_\pi + m_\ell,
\end{equation}
together with a lepton mixing $V^R_{\ell\, 2} \simeq V^R_{\ell\, 3} \approx 1$, up to $\sim1\%$, depending on the exact value of $M_{W_R}$. Since $N_1$ already couples to $\tau$  (and coupling the diluter to $\tau$ would result in a Boltzmann suppression anyway), the two remaining state couple to $e$ and $\mu$, respectively and we end up with the following mass spectrum
\begin{align}\label{flavorspectra}
	\mathbf{V}^R_\ell &\approx \begin{pmatrix}
		0 & 0 & 1
		\\
		0 & 1 & 0 
		\\
		1 & 0 & 0 
	 \end{pmatrix}, \quad \quad
	 \begin{array}{rl}
	 m_{N_1} & \sim \text{ keV},
	 \\
	 m_{N_2} & \approx m_\pi + m_\mu,
	 \\
	 m_{N_3} & \approx m_\pi + m_e.
	 \end{array}
\end{align}

\begin{figure}[t]
  \hspace{-0.2cm} \centerline{
  \includegraphics[width=9.5cm]{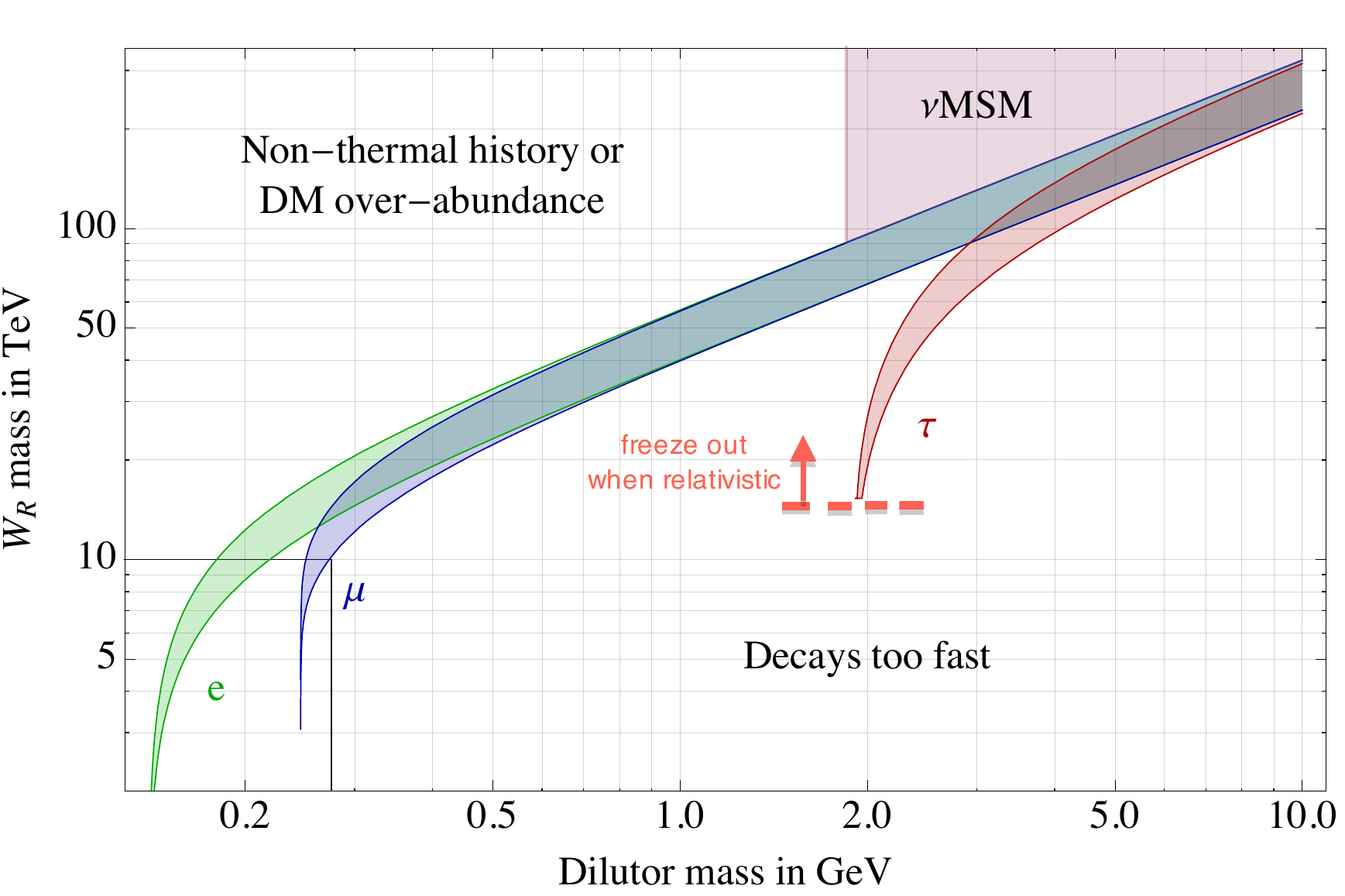}
  \raisebox{13pt}{\includegraphics[width=5.05cm]{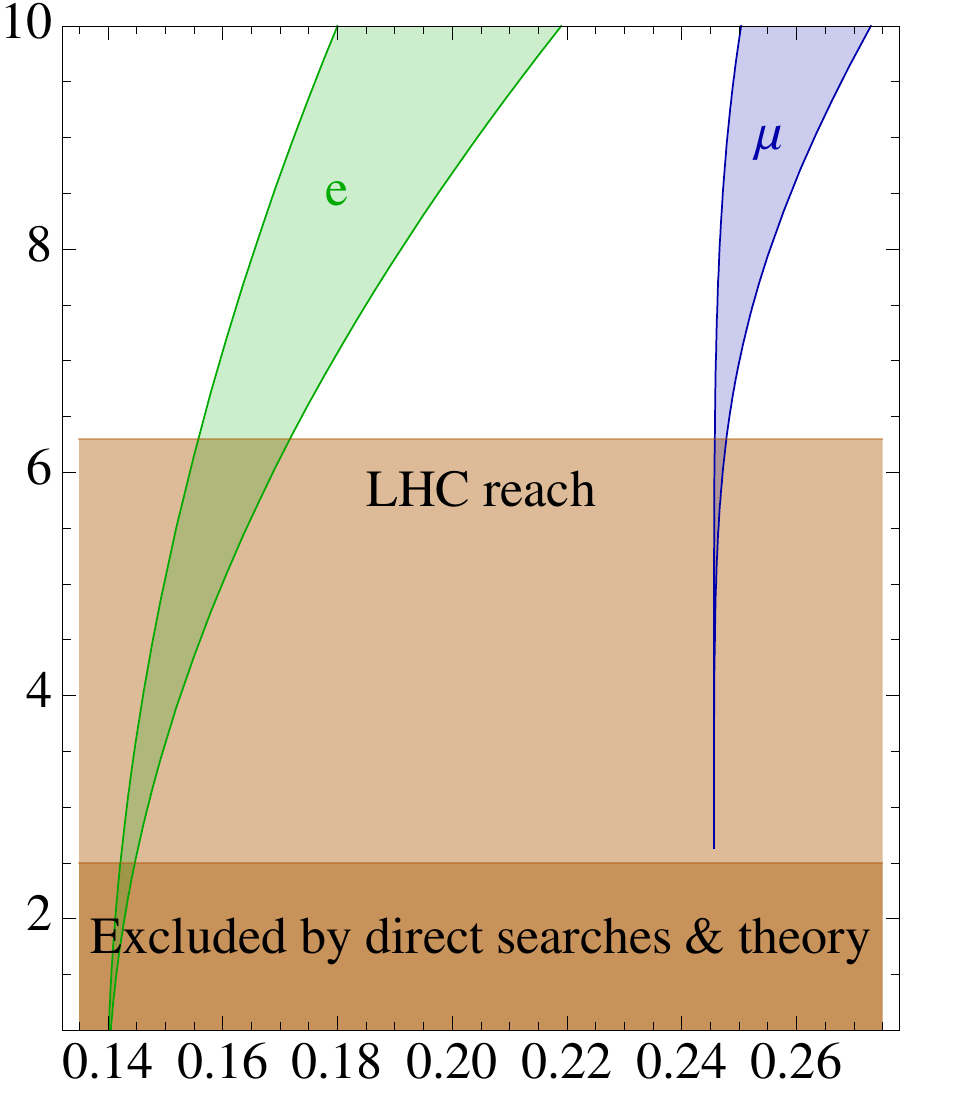}}}
  \caption{Regions of RH neutrino masses relevant for the acceptable amount of DM in the LRSM. The shaded regions (green, blue and red) labeled $e, \mu$ and $\tau$ correspond to the lifetime of the diluting $N$ between 0.5 and 2 seconds. {\bf Zoom.} Region of $M_{W_R}\lesssim10\,$TeV of our primary interest. Also shown are the theoretical lower limit from kaon mixing (which coincides with the lower limit set by the current LHC direct search), as well as the 14 TeV LHC reach.
  \label{figPhaseSpace}}
\end{figure}

Precise values depend slightly on the mass of $M_{W_R}$, as is shown in Fig.~\ref{figPhaseSpace}. The region related to the $\tau$ coupled $N$ terminates at around $M_{W_R} \simeq 15 \text{ TeV}$ due to the Boltzmann suppression. Below the shaded regions, DM is typically overproduced. Above, the decays proceed via Yukawa couplings, just as in the $\nu$MSM~\cite{Asaka:2005an, Asaka:2006ek} case, where $m_N \gtrsim 1-2 \text{ GeV}$ is obtained for a sufficient dilution, the magenta triangle in Figure~\ref{figPhaseSpace}. There is an important exception between these two scenarios, regarding the production. In the $\nu$MSM, DM is produced in a non-thermal way~\cite{Dodelson:1993je} whereas in this case, the production is thermal. The two may coincide if the reheating temperature after inflation is low - on the order of GeV.

\paragraph{Summary and a window for the LR scale}

\begin{figure}[t] \centerline{
\includegraphics[width=7.5cm]{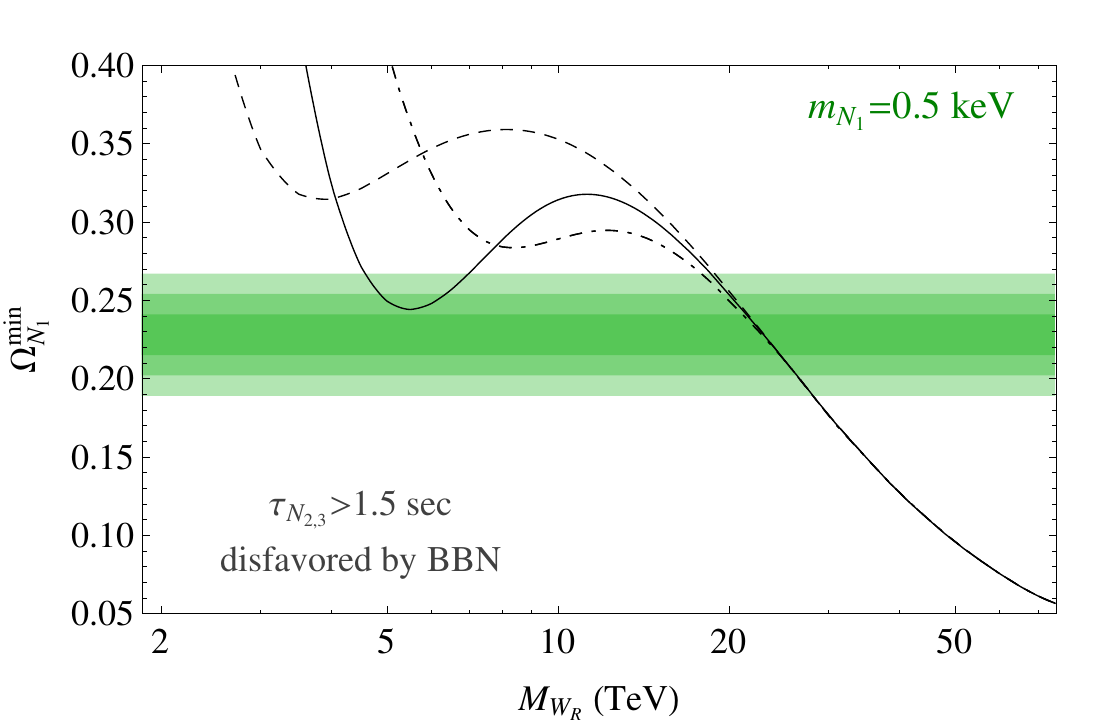}
\includegraphics[width=7.5cm]{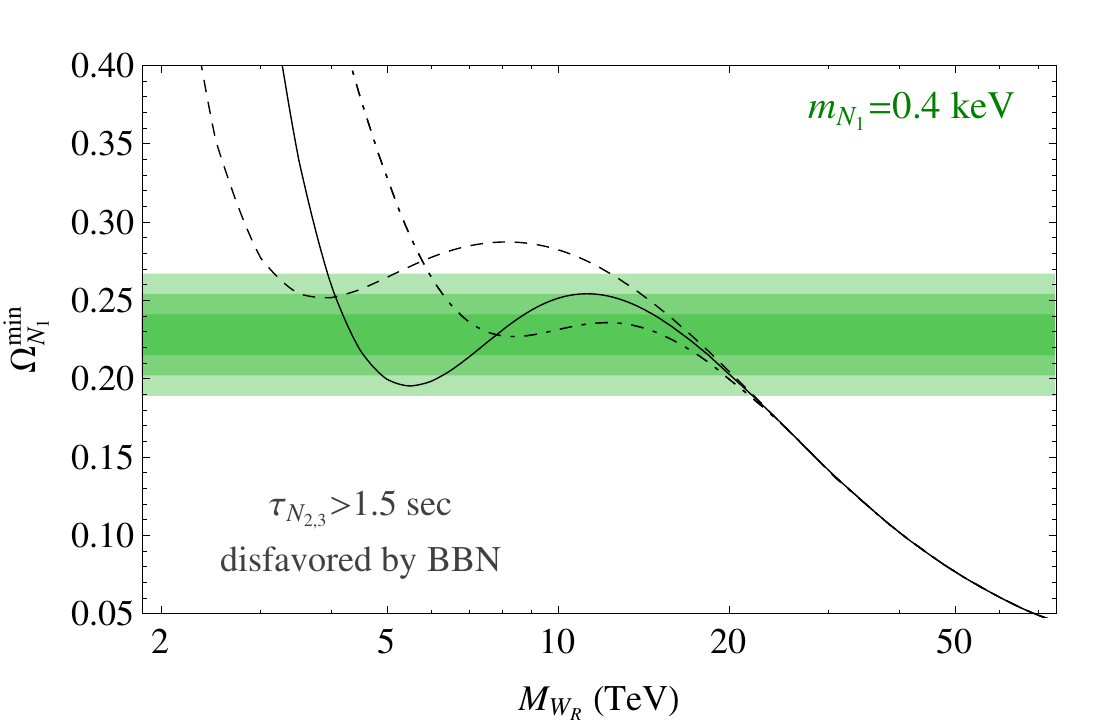}}
\caption{Minimal amount of DM relic density in the LRSM, depending on the $W_R$ mass for two fixed values of $N_1$ mass. Solid, dashed and dot-dashed lines correspond to $T_{\text{QCD}} = 350 \text{ MeV}$ (nearly first order) and $T_{\text{QCD}} = 150, 400 \text{ MeV}$ (second order), respectively and a fixed lifetime $\tau_{N_{2,3}} = 1.5 \text{ sec}$. The green bands characterise the observed DM relic abundance at 1, 2, 3\,$\sigma$ confidence level from WMAP fit.}\label{figRelicDensity}
\end{figure}

The above estimates have been improved by a detailed numerical study, described in full detail in~\cite{Nemevsek:2012cd}. This work takes into account the change of $g_*$ at different possible transition temperatures, uses exact thermally averaged rates (see appendix A and B of~\cite{Nemevsek:2012cd}) and considers a simultaneous impact of two diluters. 

The final result can be summarised succinctly in Fig.~\ref{figRelicDensity}. The measured DM relic density is shown in green bands, together with the minimal possible value of warm DM in the LRSM. There is a clearly visible dip in the obtained matter density, which is a direct result from the separation of the two freeze-out temperatures - the one of DM and that of the diluters. The position and depth of the dip depends on the exact value and nature of the QCD phase transition. The important point here is that the window opens up in the few TeV region, right within the 14 TeV LHC reach~\cite{Ferrari:2000sp}.

When mass of $W_R$ is larger than 20 TeV or so, we pass over the QCD phase transition and the freeze-out temperature for all RH neutrinos coincide, thus the window is shut. Since the relic abundance scales linearly with the mass, the window also disappears when the mass of DM is decreased. The origin of the overall falling trend of the curve for the minimal DM relic abundance seen in~\ref{figRelicDensity} is also easy to understand. It is simply because the mass of the diluter increases with the LR scale when the lifetime is kept constant and since the amount of entropy produced scales linearly with the mass of the diluter, smaller values of DM abundance can be obtained.

%
%
\paragraph{Additional constraints}

\begin{table}[t]
\centering
\begin{tabular}{cccc}
        \hline
	Constraints & $m_{N_1}$ & $\tau_{N}$ & $M_{W_R}$ \\ 
	\hline
	Dwarf Galaxy & $\gtrsim 0.4-0.5\,$keV & --- & --- 
	\\ 
	Lyman-$\alpha$ & $\gtrsim 0.5$\,--\,$1\,$keV & --- & --- 
	\\ 
	BBN \& CMB & --- & $\lesssim 1.5\,$sec & --- 
	\\ 
	$0\nu2\beta$ & --- & --- & $\gtrsim6-8\,$TeV 
	\\ 
	LHC-14 reach & $0-M_{W_R}$ & --- & $\lesssim 6.3\,$TeV 
	\\ 
	A sample point & 0.5\,keV & 1.5\,sec & $4-7$\,TeV \\
	\hline
\end{tabular}
\caption{Various constraints on the masses and lifetime of relevant states within the LRSM, coming from astrophysical, cosmological and terrestrial experiments, together with a sample point in the DM scenario. \label{WDMBound}}
\end{table}

There are important cosmological, astrophysical and low energy constraints on the LRSM warm DM scenario. They are related to the mass of DM, the lifetime of diluters and the LR scale itself and are summarized in Table~\ref{WDMBound}. A complete discussion of the limits and a list of references is presented in~\cite{Nemevsek:2012cd}, here we outline the two most stringent cases.

The most reliable bound on the DM mass is a result from the study of dwarf spheroidal galaxies. If fermionic DM inside such astrophysical objects can be regarded as a degenerate Fermi gas, a lower bound on its mass $m_{\rm DM} > 0.468^{+0.137}_{-0.082} \text{ keV}$ can be obtained~\cite{Boyarsky:2008ju}. A more sophisticated analysis which compares the maximum phase space density~\cite{Tremaine:1979we} with observations gives a slightly stronger bound, $m_{\rm DM}>0.557^{+0.163}_{-0.097} \text{ keV}$.

Another important constraint on the LR scale comes from the searches for neutrinoless double beta decay. As already emphasised above, the Majorana nature of heavy neutrinos plays an important role in the study of DM. The associated lepton number violation manifests itself at low energies and provide additional contributions to the decay rate~\cite{Mohapatra:1980yp}. Recently, an in-depth study has been performed~\cite{Tello:2010am}, which demonstrated the profound connection between $0\nu2\beta$, LNV at colliders~\cite{Keung:1983uu} and lepton flavor violation and a limit on the LR symmetry scales was derived~\cite{Nemevsek:2011aa} as a function of the signal strength, depending on the flavour structure. In our case, the mixing matrix $\mathbf{V_\ell^R}$ is diagonal and this limit becomes quite acute. From Fig.~1 of~\cite{Nemevsek:2011aa} one can conclude that the mass of $W_R$ should lie above $\sim$\,6\,--\,8\,TeV. Due to fairly large uncertainties in the calculation of nuclear matrix elements~\cite{Rodejohann:2011mu}, one cannot rule out the $W_R$ window with certainty, but it seems that an observation should be imminent and more importantly experiments are about to probe deep within this region~\cite{KamLANDZen:2012aa}.

%
%
\section{Conclusions}

The problem of overproduction of DM in the presence of super weak interactions in the early universe has been well known in the past and has prevented incorporating warm dark matter in a low scale LR symmetry. It is therefore surprising that a detailed study reveals a possible window for the LR scale in the few TeV region and at the same time ends up with predicted masses and mixings for heavy RH neutrinos. The following picture emerges: the lightest RH neutrino with a mass of around 0.5 keV coupled to $\tau$ via $W_R$ acts as warm dark matter. The remaining two RH neutrinos act as long-lived diluters and have masses $m_\pi + m_{e,\mu}$ with nearly diagonal couplings to the electron and the muon.

The resulting window is within the reach of the LHC~\cite{Ferrari:2000sp} and gives a final state with a single lepton and missing energy, which can be probed at the LHC. Low energy signals are similar to those of the $\nu$MSM searches for sterile neutrinos~\cite{Gorbunov:2007ak}. The most noteworthy exception is the predicted rate for neutrinoless double beta decay~\cite{Tello:2010am, Nemevsek:2012cd}, which is on the edge of exclusion with recent data~\cite{KamLANDZen:2012aa}. The resulting flavour composition is nearly diagonal, therefore lepton flavour violating processes are largely suppressed. Any signal from these experiments would most likely rule out this scenario.

\begin{theacknowledgments}
I would like to thank the organisers for support and a pleasant and stimulating workshop in South Dakota. I am thankful to Goran Senjanovi\'c and Yue Zhang for careful reading of the manuscript.
\end{theacknowledgments}

\bibliographystyle{aipproc}

\end{document}